\documentclass[showpacs, showkeys]{revtex4}

\usepackage[small]{caption}
\usepackage{amsmath,amsfonts,amscd,amssymb,epsf, epsfig,mathrsfs}\usepackage{graphicx}

\newcommand{\pa}{\partial}
\newcommand{\vep}{\varepsilon}
\begin{document}
\title{Ferminoic Casimir effect between spheres}
\author{L. P. Teo}
 \email{LeePeng.Teo@nottingham.edu.my}
 \affiliation{Department of Applied Mathematics, Faculty of Engineering, University of Nottingham Malaysia Campus, Jalan Broga, 43500, Semenyih, Selangor Darul Ehsan, Malaysia.}
\begin{abstract}
We consider the Casimir interaction between two spheres corresponding to  massless Dirac fields with MIT-bag boundary conditions. Using operator approach, we derive the TGTG-formula for the Casimir interaction energy between the two spheres. A byproduct is the explicit formula for the translation matrix that relates the fermionic spherical waves in different coordinate systems. In the large separation limit, it is found that the order of the Casimir interaction energy is $L^{-5}$, where $L$ is the separation between the centers of the spheres. This order is intermediate between that of two Dirichlet spheres (of order $L^{-3}$) and two Neumann spheres (of order $L^{-7}$). In the small separation limit, we derive analytically the asymptotic expansion of the Casimir interaction energy up to the next-to-leading order term. The leading term agrees with the proximity force approximation. The result for the next-to-leading order term is compared to the corresponding results for scalar fields and electromagnetic fields.
\end{abstract}
 \maketitle
\section{Introduction}

The Casimir effect induced by the vacuum energy of a quantum field is an important topic of research in various areas of theoretical physics such as quantum field theory, gravitation and cosmology, atomic physics, nanotechnology and condensed matters. In the end of the last century, the advancement in Casimir experiments have driven the acceleration of the theoretical research of Casimir effect.
In the beginning of this century, we have observed a flourish in the Casimir research into new eras. In particular, there is a breakthrough in the understanding of how to compute the Casimir interaction between two nonplanar objects using multiple scattering or mode summation approach after the hardwork of a few groups of researchers \cite{21,22,23,24,25,26,27,28,29,14,15,30,31,32,33,34,35,36,1}.

In these ten years, many research have been done to understand the Casimir interaction between two nontrivial configurations such as a sphere and a plate,  two spheres, two cylinders, a cylinder and a plate, etc. Using multiple scattering formalism, one can show that the Casimir interaction energy can be written as the integral over the trace of the logarithm of a matrix, which is the multiplication of four matrices, two of them are the T-operators  of the two objects, and another two are the translation matrices that relate the coordinate systems used for the two objects. The two T-operators are intimately related to the scattering matrices of the objects and can be easily computed by matching boundary conditions. The  difficult part of the problem is to derive the translation matrices, and one usually needs advanced mathematical tools.

In the pioneering work \cite{37, 6}, Milton has considered the fermionic Casimir effect of a spherical bag. Subsequently, Casimir effect of fermionic fields has aroused considerable interest \cite{38,5,39,40,41,42,43,44,45,46,47,48,49,50,51,52,53,54,55,56}. Nevertheless, to the best of our knowledge, no one has considered the fermionic Casimir interaction between two nontrivial objects. In this work, we take the first step in this direction. We consider the fermionic Casimir effect between two spheres exterior to each other and compare the results to that of scalar fields and electromagnetic fields. The fermionic field satisfies the Dirac equation and is subject to the MIT-bag boundary conditions. As has been well-known, although fermionic fields have spin and statistics that are different from that of bosonic fields, they also induced attractive Casimir force on two parallel plates. Hence, it is natural to expect the same result in the case of two spheres. We show that this is indeed the case by computing the large separation and small separation asymptotic behaviors.

\section{The Casimir energy}

Consider two spheres $A$ and $B$ outside each other with radii $r_A$ and $r_B$ respectively. Let $L$ be the distance between the centers of the spheres and let $d$ be the distance between the spheres. Then $d=L-r_A-r_B$.

We want to study the Casimir effect due to the vacuum fluctuations of a massless spin $1/2$- fermionic field with MIT-bag boundary conditions on the spheres.
The equation of motion   is the Dirac equation
\begin{equation*}
i\gamma^{\mu}\nabla_{\mu}\psi=0,
\end{equation*}
where
$
\nabla_{\mu}=\pa_{\mu}+\Gamma_{\mu},
$ and $\Gamma_{\mu}$ is the spin connection. On the boundaries of the spheres, MIT-bag boundary conditions are imposed:
\begin{equation}\label{eq5_16_2}
(1+i\gamma^{\mu}n_{\mu})\psi\Bigr|_{\text{boundary}}=0,
\end{equation}where $n_{\mu}$ is the unit outward normal vector.

 To derive the formula for the Casimir interaction energy between the spheres, we use the formalism we developed in \cite{1}. The TGTG formula for the Casimir interaction energy is given by
\begin{equation}\label{eq02_09_1}
E_{\text{Cas}}=-\frac{\hbar  }{2\pi}\int_0^{\infty} d\xi \ln\det\left(\mathbb{I}-\mathbb{T}^A\mathbb{G}^{AB}\mathbb{T}^B\mathbb{G}^{BA}\right).
\end{equation}Here $\xi$ is the imaginary frequency. The minus sign in front of the integral appears because we are dealing with fermionic fields. In this formula, $\mathbb{T}^A$ and  $\mathbb{T}^B$ are respectively the T-operators of the spheres $A$ and $B$, $\mathbb{G}^{AB}$ and $\mathbb{G}^{BA}$ are the translation matrices that relate the bases of wave functions in different coordinate systems.

In spherical coordinates, the wave basis are parametrized by $(j, m)$ with $j=1/2, 3/2, 5/2, \ldots$ and $m=-j, -j+1, \ldots, j-1, j$. They are given by
\begin{equation*}
\begin{split}
\psi_{jm,1}^{(\pm), *}=&\mathcal{C}_j^{*} e^{\mp i\omega t}  \begin{pmatrix}f^*_{j-\frac{1}{2}}(kr)\Omega_{j,j-\frac{1}{2},m}\\
\mp i f^*_{j+\frac{1}{2}}(kr)\Omega_{j,j+\frac{1}{2},m}\end{pmatrix},\\
\psi_{jm,2}^{(\pm), *}=&\mathcal{C}_j^{*}e^{\mp i\omega t} \begin{pmatrix}f^*_{j+\frac{1}{2}}(kr)\Omega_{j,j+\frac{1}{2},m}\\
\pm i f^*_{j-\frac{1}{2}}(kr)\Omega_{j,j-\frac{1}{2},m}\end{pmatrix},
\end{split}
\end{equation*}
where $k=\omega/c$, $\psi^{(+)}$ and $\psi^{(-)}$ are respectively the positive energy modes and negative energy modes, $*=$ reg or out for regular or outgoing waves,
\begin{equation*}\begin{split}
f^{\text{reg}}_l(z)=&j_l(z)=\sqrt{\frac{\pi}{2z}}J_{l+\frac{1}{2}}(z),\\
 f^{\text{out}}_l(z)=&h_l^{(1)}(z)=\sqrt{\frac{\pi}{2z}}H^{(1)}_{l+\frac{1}{2}}(z);\end{split}
\end{equation*}  $\Omega_{jlm}$ are the spherical harmonic spinors \cite{3}:
\begin{equation*}
\begin{split}
\Omega_{jlm}=&\left(\begin{aligned} \sqrt{\frac{j+m}{2j}}Y_{l,m-\frac{1}{2}}\\\sqrt{\frac{j-m}{2j}}Y_{l,m+\frac{1}{2}}\end{aligned}\right),\hspace{1cm}j=l+\frac{1}{2},\\
\Omega_{jlm}=&\left(\begin{aligned}- \sqrt{\frac{j-m+1}{2j+2}}Y_{l,m-\frac{1}{2}}\\\sqrt{\frac{j+m+1}{2j+2}}Y_{l,m+\frac{1}{2}}\end{aligned}\right),\hspace{1cm}j=l-\frac{1}{2}
\end{split}
\end{equation*}
and the constants
\begin{equation*}
 \mathcal{C}_j^{\text{reg}}=i^{-j+\frac{1}{2}},\quad \mathcal{C}_j^{\text{out}}=\frac{\pi}{2} i^{j+\frac{3}{2}}
 \end{equation*} are  introduced to facilitate the change to imaginary frequencies.

Matching the boundary conditions on the spheres, we find that the T-operator for sphere $A$ is given by
\begin{align*}
\mathbb{T}_{jm}^{(\pm),A}=\begin{pmatrix} T_{jm,1}^{(\pm), A}&0\\0&T_{jm, 2}^{(\pm), A}\end{pmatrix}
\end{align*}with
\begin{align*}
T_{jm,1}^{(\pm), A}=\frac{I_j(\kappa r_A)\mp iI_{j+1}(\kappa r_A)}{K_j(\kappa r_A)\pm iK_{j+1}(\kappa r_A)},\\
T_{jm,2}^{(\pm), A}=\frac{I_j(\kappa r_A)\pm iI_{j+1}(\kappa r_A)}{K_j(\kappa r_A)\mp iK_{j+1}(\kappa r_A)},
\end{align*}and $k=i\kappa$. The T-operator for sphere $B$ -- $\mathbb{T}^B$ is given by the same formula by changing $r_A$ to $r_B$.

The derivation of the $\mathbb{G}^{AB}$ and $\mathbb{G}^{BA}$ matrices are more complicated.   Their components   are defined by the relations
\begin{equation}\label{eq3_3_3}\begin{split}
\begin{pmatrix}\psi^{(\pm), \text{out}}_{j'm', 1 }(\mathbf{x}',\omega)\\\psi^{(\pm), \text{out}}_{j'm', 2 }(\mathbf{x}',\omega)\end{pmatrix} =&\sum_{j }\sum_{m} \mathbb{G}^{AB, (\pm)}_{jm, j'm'}\begin{pmatrix}\psi_{jm,1}^{(\pm),\text{reg}}(\mathbf{x},\omega)\\ \psi_{jm,2}^{(\pm),\text{reg}}(\mathbf{x},\omega)\end{pmatrix},\\
\begin{pmatrix} \psi^{(\pm), \text{out}}_{jm, 1 }(\mathbf{x},\omega)\\\psi^{(\pm), \text{out}}_{jm, 2 }(\mathbf{x},\omega)\end{pmatrix} =&\sum_{j' }\sum_{m'} \mathbb{G}^{BA, (\pm)}_{j'm', jm}\begin{pmatrix}\psi_{j'm',1}^{(\pm),\text{reg}}(\mathbf{x}',\omega)\\ \psi_{j'm',2}^{(\pm),\text{reg}}(\mathbf{x}',\omega)\end{pmatrix},
\end{split}\end{equation}where $\mathbf{x}'=\mathbf{x}-\mathbf{L}$, $\mathbf{L}=(0,0,L)$.

To compute these matrices explicitly, let us use the operator method introduced in \cite{4} and developed in \cite{1}.
As in \cite{4,1}, define a differential operator $\mathcal{P}_{lm}$ by
\begin{equation*}
\begin{split}
\mathcal{P}_{lm}=&(-1)^m \sqrt{\frac{2l+1}{4\pi}\frac{(l-m)!}{(l+m)!}}\left(\frac{\pa_x+i\pa_y}{ik}\right)^mP_l^{(m)}\left(\frac{\pa_z}{ik}\right), \\
\mathcal{P}_{l,-m}=&  \sqrt{\frac{2l+1}{4\pi}\frac{(l-m)!}{(l+m)!}}\left(\frac{\pa_x-i\pa_y}{ik}\right)^mP_l^{(m)}\left(\frac{\pa_z}{ik}\right).
\end{split}
\end{equation*}Here $P_l^{(m)}(z)$ is the $m$ times derivative of the Legendre polynomial $P_l(z)$.
This operator is such that
\begin{equation*}
\mathcal{P}_{lm}e^{i\mathbf{k}\cdot\mathbf{r}}=Y_{lm}(\theta_k,\phi_k)e^{i\mathbf{k}\cdot\mathbf{r}},
\end{equation*}where $$\mathbf{k}=(k_1,k_2,k_3)=(k\sin\theta_k\cos\phi_k, k\sin\theta_k\sin\phi_k, k\cos\theta_k),$$ and $$\mathbf{r}=(x,y,z)=(r\sin\theta\cos\phi, r\sin\theta\sin\phi, r\cos\theta).$$
It has been shown in \cite{1,4} that
\begin{equation}\label{eq3_3_1}\begin{split}
\mathcal{P}_{lm}j_0(kr)=&i^lj_l(kr)Y_{lm}(\theta,\phi),\\
\mathcal{P}_{lm}h_0^{(1)}(kr)=&i^lh_l^{(1)}(kr)Y_{lm}(\theta,\phi);
\end{split}\end{equation}and
\begin{equation}\label{eq3_3_2}\begin{split}
j_0(kr)=&\frac{1}{4\pi}\int_0^{2\pi}d\phi_k\int_0^{\pi}d\theta_k\,\sin\theta_k e^{i\mathbf{k}\cdot\mathbf{r}}\\
h_0^{(1)}(kr)=&\frac{1}{2\pi}\int_{-\infty}^{\infty}dk_x\int_{-\infty}^{\infty}dk_y\frac{e^{ik_xx+ik_yy\pm i\sqrt{k^2-k_x^2-k_y^2}z}}{k\sqrt{k^2-k_x^2-k_y^2}},\quad z\gtrless 0.
\end{split}\end{equation}From \eqref{eq3_3_1}, we have
\begin{align*}
\psi_{jm,1}^{(\pm), *}=&\mathcal{C}_j^{*} e^{\mp i\omega t} i^{-j+\frac{1}{2}}\left(\begin{aligned} \sqrt{\frac{j+m}{2j}}\mathcal{P}_{j-\frac{1}{2},m-\frac{1}{2}}\\\sqrt{\frac{j-m}{2j}}\mathcal{P}_{j-\frac{1}{2},m+\frac{1}{2}}\\
\pm\sqrt{\frac{j-m+1}{2j+2}}\mathcal{P}_{j+\frac{1}{2},m-\frac{1}{2}}\\\mp\sqrt{\frac{j+m+1}{2j+2}}\mathcal{P}_{j+\frac{1}{2},m+\frac{1}{2}}\end{aligned}\right)f^*_0(kr)\\
\psi_{jm,2}^{(\pm), *}=&-\mathcal{C}_j^{*} e^{\mp i\omega t} i^{-j-\frac{1}{2}}\left(\begin{aligned}
\sqrt{\frac{j-m+1}{2j+2}}\mathcal{P}_{j+\frac{1}{2},m-\frac{1}{2}}\\-\sqrt{\frac{j+m+1}{2j+2}}\mathcal{P}_{j+\frac{1}{2},m+\frac{1}{2}}\\
\pm\sqrt{\frac{j+m}{2j}}\mathcal{P}_{j-\frac{1}{2},m-\frac{1}{2}}\\\pm\sqrt{\frac{j-m}{2j}}\mathcal{P}_{j-\frac{1}{2},m+\frac{1}{2}}\end{aligned}\right)f_0^*(kr)
\end{align*}
Now, define
\begin{align*}
\boldsymbol{\mathcal{P}}_{jm,1}^{\pm}=&\frac{(-1)^{m-\frac{1}{2}}}{2}\left(\begin{aligned} \sqrt{\frac{j+m}{2j}}\mathcal{P}_{j-\frac{1}{2},-m+\frac{1}{2}}\\-\sqrt{\frac{j-m}{2j}}\mathcal{P}_{j-\frac{1}{2},-m-\frac{1}{2}}\\
\pm\sqrt{\frac{j-m+1}{2j+2}}\mathcal{P}_{j+\frac{1}{2},-m+\frac{1}{2}}\\\pm\sqrt{\frac{j+m+1}{2j+2}}\mathcal{P}_{j+\frac{1}{2},-m-\frac{1}{2}}\end{aligned}\right),\\
\boldsymbol{\mathcal{P}}_{jm,2}^{\pm}=&\frac{(-1)^{m-\frac{1}{2}}}{2}\left(\begin{aligned} \sqrt{\frac{j-m+1}{2j+2}}\mathcal{P}_{j+\frac{1}{2},-m+\frac{1}{2}}\\\sqrt{\frac{j+m+1}{2j+2}}\mathcal{P}_{j+\frac{1}{2},-m-\frac{1}{2}}\\ \pm\sqrt{\frac{j+m}{2j}}\mathcal{P}_{j-\frac{1}{2},-m+\frac{1}{2}}\\\mp\sqrt{\frac{j-m}{2j}}\mathcal{P}_{j-\frac{1}{2},-m-\frac{1}{2}}\end{aligned}\right).
\end{align*}Using the orthogonality of the spherical harmonics and the formulas \eqref{eq3_3_2}, one can check directly  that
\begin{align*}
\boldsymbol{\mathcal{P}}_{j'm',1}^{\pm}\cdot \psi_{jm,1}^{(\pm), \text{reg}}(\mathbf{x},\omega)\Biggr|_{\mathbf{x}=0}=&(-1)^{-j+\frac{1}{2}}\frac{e^{\mp i\omega t}}{4\pi}\delta_{jj'}\delta_{mm'},\\
\boldsymbol{\mathcal{P}}_{j'm',2}^{\pm}\cdot \psi_{jm,1}^{(\pm), \text{reg}}(\mathbf{x},\omega)\Biggr|_{\mathbf{x}=0}=&0,\\
\boldsymbol{\mathcal{P}}_{j'm',1}^{\pm}\cdot \psi_{jm,2}^{(\pm), \text{reg}}(\mathbf{x},\omega)\Biggr|_{\mathbf{x}=0}=&0\\
\boldsymbol{\mathcal{P}}_{j'm',2}^{\pm}\cdot \psi_{jm,2}^{(\pm), \text{reg}}(\mathbf{x},\omega)\Biggr|_{\mathbf{x}=0}=&(-1)^{-j+\frac{1}{2}}i\frac{e^{\mp i\omega t}}{4\pi}\delta_{jj'}\delta_{mm'}.
\end{align*}From these and the definition \eqref{eq3_3_3}, we find that
\begin{align*}
\mathbb{G}^{AB, (\pm)}_{jm, j'm'}=&(-1)^{j-\frac{1}{2}}4\pi e^{\pm i\omega t} \begin{pmatrix}
\boldsymbol{\mathcal{P}}_{jm,1}^{\pm}\cdot\psi^{(\pm), \text{out}}_{j'm', 1 }(\mathbf{x}',\omega)\bigr|_{\mathbf{x}=0}
&-i\boldsymbol{\mathcal{P}}_{jm,2}^{\pm}\cdot\psi^{(\pm), \text{out}}_{j'm', 1 }(\mathbf{x}',\omega)\bigr|_{\mathbf{x}=0}\\
 \boldsymbol{\mathcal{P}}_{jm,1}^{\pm}\cdot\psi^{(\pm), \text{out}}_{j'm', 2 }(\mathbf{x}',\omega)\bigr|_{\mathbf{x}=0}&-i\boldsymbol{\mathcal{P}}_{jm,2}^{\pm}\cdot\psi^{(\pm), \text{out}}_{j'm', 2 }(\mathbf{x}',\omega)\bigr|_{\mathbf{x}=0}\end{pmatrix}.
\end{align*} After some computations, we obtain the following formulas for $\mathbb{G}^{AB, (\pm)}_{jm, j'm'}$ and $ \mathbb{G}^{BA, (\pm)}_{j'm', jm}$:
\begin{align*}
\mathbb{G}^{AB, (\pm)}_{jm, j'm'}=&(-1)^{j'+m}\delta_{m,m'}\frac{\pi }{2}\sqrt{\frac{(j-m)!(j'-m')!}{(j+m)!(j'+m')!}}\int_{0}^{\infty} d\theta\sinh\theta
 e^{ -\kappa L\cosh\theta} \\
&\times \begin{pmatrix}(j+m)P_{j-\frac{1}{2}}^{m-\frac{1}{2}}(\cosh\theta) &  P_{j-\frac{1}{2}}^{m+\frac{1}{2}}(\cosh\theta)\\
-i(j-m+1)P_{j+\frac{1}{2}}^{m-\frac{1}{2}}(\cosh\theta)&iP_{j+\frac{1}{2}}^{m+\frac{1}{2}}(\cosh\theta) \end{pmatrix}\begin{pmatrix}
(j'+m')
 P_{j'-\frac{1}{2}}^{m'-\frac{1}{2}}(\cosh\theta)& i(j'-m'+1)P_{j'+\frac{1}{2}}^{m'-\frac{1}{2}}(\cosh\theta)\\-P_{j'-\frac{1}{2}}^{m'+\frac{1}{2}}(\cosh\theta)
 &iP_{j'+\frac{1}{2}}^{m'+\frac{1}{2}}(\cosh\theta)\end{pmatrix},
\end{align*}and
\begin{equation*}
 \mathbb{G}^{BA, (\pm)}_{j'm', jm}=\begin{pmatrix} 1 & 0\\0&-1\end{pmatrix} \mathbb{G}^{AB, (\pm)}_{jm, j'm'} \begin{pmatrix} 1 & 0\\0&-1\end{pmatrix}.
\end{equation*}
Hence, after some simplification, the Casimir interaction energy between two fermionic spheres  can be written as  \begin{equation}\label{eq02_12_1}
E_{\text{Cas}}=-\frac{\hbar  }{\pi}\int_0^{\infty} d\xi \ln\det\left(\mathbb{I}-\mathbb{M}\right),
\end{equation}where
\begin{align*}
\mathbb{M}_{jm,j'm'}=&\frac{\pi^2}{4}\delta_{m,m'} \begin{pmatrix} T_{jm,1}^{+, A}&0\\0&T_{jm, 2}^{+, A}\end{pmatrix}
 \sum_{j''}\mathbb{V}^1_{jm,j''m}
 \begin{pmatrix} T_{j''m,1}^{+, B}&0\\0&T_{j''m, 2}^{+, B}\end{pmatrix}\mathbb{V}^2_{j''m,j'm},
\end{align*}with
\begin{align*}
\mathbb{V}^{1}_{jm, j''m}=& \sqrt{\frac{(j-m)!(j''-m)!}{(j+m)!(j''+m)!}} \int_{0}^{\infty} d\theta\sinh\theta
 e^{ -\kappa L\cosh\theta} \\
&\times \begin{pmatrix}(j+m)P_{j-\frac{1}{2}}^{m-\frac{1}{2}}(\cosh\theta) &  P_{j-\frac{1}{2}}^{m+\frac{1}{2}}(\cosh\theta)\\
-(j-m+1)P_{j+\frac{1}{2}}^{m-\frac{1}{2}}(\cosh\theta)&P_{j+\frac{1}{2}}^{m+\frac{1}{2}}(\cosh\theta) \end{pmatrix}\begin{pmatrix}
(j''+m)
 P_{j''-\frac{1}{2}}^{m-\frac{1}{2}}(\cosh\theta)& (j''-m+1)P_{j''+\frac{1}{2}}^{m-\frac{1}{2}}(\cosh\theta)\\-P_{j''-\frac{1}{2}}^{m+\frac{1}{2}}(\cosh\theta)
 &P_{j''+\frac{1}{2}}^{m+\frac{1}{2}}(\cosh\theta)\end{pmatrix},
\end{align*}and
\begin{align*}
\mathbb{V}^{2}_{j''m, j'm}= & \mathbb{V}^{1}_{j'm, j''m}.
\end{align*}In \eqref{eq02_12_1}, we have multiplied a factor of 2 taking into account the contribution from positive and negative energy modes. Although the components of the $T$-operators of the spheres $\mathbb{T}^A$ and $\mathbb{T}^B$ are complex, one can check directly that after taking the trace, the contribution to the Casimir interaction energy from positive energy modes and negative energy modes are both real and equal to each other.

We would also like to remark that using the identities of associated Legendre functions, one can show that the product
\begin{align*}
\begin{pmatrix}(j+m)P_{j-\frac{1}{2}}^{m-\frac{1}{2}}(\cosh\theta) &  P_{j-\frac{1}{2}}^{m+\frac{1}{2}}(\cosh\theta)\\
-(j-m+1)P_{j+\frac{1}{2}}^{m-\frac{1}{2}}(\cosh\theta)&P_{j+\frac{1}{2}}^{m+\frac{1}{2}}(\cosh\theta) \end{pmatrix}\begin{pmatrix}
(j''+m)
 P_{j''-\frac{1}{2}}^{m-\frac{1}{2}}(\cosh\theta)& (j''-m+1)P_{j''+\frac{1}{2}}^{m-\frac{1}{2}}(\cosh\theta)\\-P_{j''-\frac{1}{2}}^{m+\frac{1}{2}}(\cosh\theta)
 &P_{j''+\frac{1}{2}}^{m+\frac{1}{2}}(\cosh\theta)\end{pmatrix}
\end{align*}is a matrix of the form
\begin{align*}
\begin{pmatrix} A & B\\-B & -A\end{pmatrix}.
\end{align*}In fact, $\mathbb{V}^{1}_{jm, j'm}$ can be expressed as linear combinations of $K_{j''}(\kappa L)$ with coefficients depending on $j, j', j''$ and $m$. However, it would not help our computations later.

In the following, we are going to explore the asymptotic behavior of the Casimir interaction energy when the separation between the spheres is large and when the separation is small.

\section{The large separation asymptotic behavior}
When the separation between the spheres is large, i.e., when $L\gg r_A, r_B$, the leading contribution to the Casimir interaction energy comes from terms with lowest $j$ and $m$, namely with $j=1/2$ and $m=\pm 1/2$. Replacing $\kappa$ by $\kappa/L$, we find that the leading contribution is
\begin{align}\label{eq2_27_1}
E_{\text{Cas}}\sim &\frac{2\hbar c}{\pi L} \int_0^{\infty} d\kappa\;\;\text{tr}\, \mathbb{M}_0
\end{align}
where
\begin{align*}
\mathbb{M}_0=&\frac{\pi^2}{4}\begin{pmatrix} T_{1}^A&0\\0 &T_{2}^A\end{pmatrix}
\mathbb{V}^{1}_0\begin{pmatrix} T_{1}^B &0\\0 &T_{2}^B \end{pmatrix}\mathbb{V}^{2}_0.
\end{align*}The factor $2$ in front of the integral in \eqref{eq2_27_1} comes from $m=1/2$ and $m=-1/2$ which give equal contributions. When $L\gg r_A, r_B$,
\begin{align*}
T_1^A=\overline{T_2^A}=\frac{\displaystyle I_{\frac{1}{2}}\left(\frac{\kappa r_A}{L}\right)- iI_{\frac{3}{2}}\left(\frac{\kappa r_A}{L}\right)}{\displaystyle K_{\frac{1}{2}}\left(\frac{\kappa r_A}{L}\right)+ iK_{\frac{3}{2}}\left(\frac{\kappa r_A}{L}\right)}\sim & - \frac{2i}{\pi}\left(\frac{\kappa r_A}{L}\right)^2,
\end{align*}and a similar expression for $T_1^B$ and $T_2^B$.
\begin{align*}
\mathbb{V}_0^1=\mathbb{V}_0^2
=&\int_0^{\infty} d\theta\sinh\theta e^{-\kappa\cosh\theta}\begin{pmatrix} 1 &\cosh\theta\\
-\cosh\theta&-1\end{pmatrix}\\
=&\frac{1}{\kappa^2}\begin{pmatrix}\kappa &\kappa+1\\
-(\kappa+1) & -\kappa \end{pmatrix}e^{-\kappa}.
\end{align*}
Hence,
\begin{align*}
E_{\text{Cas}}\sim &-\frac{2\hbar c r_A^2 r_B^2}{\pi L^5} \int_0^{\infty} d\kappa\, \text{tr}\;\;\begin{pmatrix} 1 & 0\\0 &-1\end{pmatrix}
\begin{pmatrix}\kappa &\kappa+1\\
-(\kappa+1) & -\kappa \end{pmatrix}\begin{pmatrix} 1 & 0\\0 &-1\end{pmatrix}
\begin{pmatrix}\kappa &\kappa+1\\
-(\kappa+1) & -\kappa \end{pmatrix}e^{-2\kappa}\\
=&-\frac{4\hbar c r_A^2 r_B^2}{\pi L^5}\int_0^{\infty} d\kappa \; (2\kappa^2+2\kappa+1)e^{-2\kappa}\\
=&-\frac{6\hbar c r_A^2 r_B^2}{\pi L^5}.
\end{align*}Namely, in the large separation regime, the order of the Casimir interaction energy is $L^{-5}$.

For two Dirichlet spheres, the large distance asymptotic behavior of the Casimir interaction energy is
\begin{equation*}
E_{\text{Cas}}^{\text{D}}\sim  -\frac{\hbar c r_Ar_B}{4\pi L^3},
\end{equation*}whereas for two Neumann spheres,
\begin{equation*}
E_{\text{Cas}}^{\text{N}}\sim  -\frac{161\hbar c r_A^3r_B^3}{96\pi L^7},
\end{equation*}
and for two perfectly conducting spheres,
\begin{equation*}
E_{\text{Cas}}^{\text{C}}\sim  -\frac{143\hbar c r_A^3r_B^3}{16\pi L^7}.
\end{equation*}
Hence, we see that in the large separation regime, the order of the Casimir interaction is intermediate between the Dirichlet and the Neumann case.

\section{The small separation asymptotic behavior}
The derivation of the small separation asymptotic behavior is more complicated. Based on the pioneering work \cite{14}, a perturbative machinery has been developed in \cite{15,7,8,9,10,11,12,13,16,17,18}. We use the same method with necessary modification to the present scenario.

First we expand the logarithm in \eqref{eq02_12_1} and get
  \begin{equation*}
E_{\text{Cas}}=\frac{\hbar c  }{\pi}\sum_{s=0}^{\infty}\frac{1}{s+1}\int_0^{\infty} d\kappa \sum_{m=\pm \frac{1}{2}, \pm \frac{3}{2}, \ldots }
\sum_{j_0=|m|, |m|+\frac{1}{2},\ldots}\ldots \sum_{j_s=|m|, |m|+\frac{1}{2},\ldots}\mathbb{M}_{j_0m,j_1m}\ldots\mathbb{M}_{j_{s}m,j_0m},
\end{equation*}
with
\begin{align*}
\mathbb{M}_{j_im,j_{i+1}m}=&\frac{\pi^2}{4}  \begin{pmatrix} T_{j_im,1}^{+, A}&0\\0&T_{j_im, 2}^{+, A}\end{pmatrix}
 \sum_{j_i'}\mathbb{V}^1_{j_im,j_i'm}
 \begin{pmatrix} T_{j_i'm,1}^{+, B}&0\\0&T_{j_i'm, 2}^{+, B}\end{pmatrix}\mathbb{V}^2_{j_i'm,j_{i+1}m}.
\end{align*}
Making a change of variables
\begin{gather*}
\vep=\frac{d}{r_A+r_B}, \quad a=\frac{r_A}{r_A+r_B},\quad b=\frac{r_B}{r_A+r_B},\quad\omega = \kappa(r_A+r_B),
\\
j_0=l,\quad j_i=l+n_i\\
j_i'=\frac{b}{2a}(l_i+l_{i+1})+q_i=\frac{b}{a}l+\frac{b}{2a}(n_i+n_{i+1})+q_i,\\
\omega=\frac{l\sqrt{1-\tau^2}}{a\tau},
\end{gather*}and replacing summations by appropriate integrations, we have
 \begin{equation}\label{eq02_12_2}
E_{\text{Cas}}\sim \frac{\hbar c  }{\pi r_A}\sum_{s=0}^{\infty}\frac{1}{s+1}\int_0^{1} \frac{d\tau}{\tau^2\sqrt{1-\tau^2}} \int_0^{\infty} dl \,l
\int_{-\infty}^{\infty} dm \int_{-\infty}^{\infty} dn_1\ldots\int_{-\infty}^{\infty} dn_s\prod_{i=0}^s\mathbb{M}_{(l+n_i)m,(l+n_{i+1})m}.
\end{equation}

Next we expand each term in small $\vep$ keeping in mind that $l\sim \vep^{-1}$, $n_i, q_i\sim \vep^{-\frac{1}{2}}$, $\tau\sim \vep^0$.
Using the formula
\begin{align*}
P_l^m(\cosh\theta)=&\frac{(l+m)!}{\pi l!}\int_0^{\pi}d\varphi \left(\cosh\theta+\sinh\theta\cos\varphi\right)^{l}\cos m\varphi\\
=&\frac{(l+m)!}{\pi  }\sum_{k=0}^l\frac{1}{k!(l-k)!}e^{(l-2k)\theta}\int_{-\frac{\pi}{2}}^{\frac{\pi}{2}}d\varphi \cos^{2l-2k}\varphi \sin^{2k}\varphi e^{2im\varphi},
\end{align*}
we find that
\begin{align*}
&\begin{pmatrix}(j_i+m)P_{j_i-\frac{1}{2}}^{m-\frac{1}{2}}(\cosh\theta) &  P_{j_i-\frac{1}{2}}^{m+\frac{1}{2}}(\cosh\theta)\\
-(j_i-m+1)P_{j_i+\frac{1}{2}}^{m-\frac{1}{2}}(\cosh\theta)&P_{j_i+\frac{1}{2}}^{m+\frac{1}{2}}(\cosh\theta) \end{pmatrix}\begin{pmatrix}
(j_i'+m)
 P_{j_i'-\frac{1}{2}}^{m-\frac{1}{2}}(\cosh\theta)& (j_i'-m+1)P_{j_i'+\frac{1}{2}}^{m-\frac{1}{2}}(\cosh\theta)\\-P_{j_i'-\frac{1}{2}}^{m+\frac{1}{2}}(\cosh\theta)
 &P_{j_i'+\frac{1}{2}}^{m+\frac{1}{2}}(\cosh\theta)\end{pmatrix}\\
 \sim &\frac{(j_i+m)!(j_i'+m)!}{\pi^2}\sum_{k=0}^{\infty}\frac{1}{k!\left(j_i+\frac{1}{2}-k\right)!}
  \sum_{k'=0}^{\infty}\frac{1}{\displaystyle k'!\left(j_i'+\frac{1}{2}-k'\right)!}\\&\times\int_{-\frac{\pi}{2}}^{\frac{\pi}{2}}d\varphi \int_{-\frac{\pi}{2}}^{\frac{\pi}{2}}d\varphi' \cos^{2j_i-2k}\varphi \sin^{2k}\varphi e^{2im\varphi}
  \cos^{2j_i'-2k'}\varphi' \sin^{2k'}\varphi' e^{2im\varphi'}\\&
  \times \exp\left(\left[\frac{l}{a}+\frac{a+1}{2a}n_i+\frac{b}{2a}n_{i+1}+q_i-2k-2k'\right]\theta\right) \begin{pmatrix} A & B\\C&D\end{pmatrix},
\end{align*}
where
\begin{align*}
A=&\frac{1}{\cos\varphi\cos\varphi'}e^{-\theta}\left(j_i+\frac{1}{2}-k\right)\left(j_i'+\frac{1}{2}-k'\right)\left( e^{-i(\varphi+\varphi')}-e^{i(\varphi+\varphi')}
\right),\\
B=&\left(j_i+\frac{1}{2}-k\right)\frac{\cos\varphi'}{\cos\varphi}\left[\left(j_i'-m+1\right)e^{-i(\varphi+\varphi')}
+\left(j_i'+m+1\right)e^{i(\varphi+\varphi')}\right],\\
C=&-\left(j_i'+\frac{1}{2}-k'\right)\frac{\cos\varphi}{\cos\varphi'}\left[\left(j_i-m+1\right)e^{-i(\varphi+\varphi')}+\left(j_{i}+m+1\right)e^{i(\varphi+\varphi')}\right],\\
D=&-e^{\theta}\cos\varphi\cos\varphi'\left[\left(j_i-m+1\right)\left(j_i'-m+1\right)e^{-i(\varphi+\varphi')}
 -\left(j_i+m+1\right)\left(j_i'+m+1\right)e^{i(\varphi+\varphi')}\right].
\end{align*}
Making a change of variables
\begin{align*}
\varphi=\frac{\widetilde{\varphi}}{\sqrt{l}},\quad \varphi'=\frac{\widetilde{\varphi}'}{\sqrt{l}},
\end{align*}we have
\begin{align*}
&\int_{-\frac{\pi}{2}}^{\frac{\pi}{2}}d\varphi  \cos^{2j_i-2k}\varphi \sin^{2k}\varphi e^{2im\varphi}\\
\sim &\frac{1}{l^{k+\frac{1}{2}}}\int_{-\infty}^{\infty} d\widetilde{\varphi} \widetilde{\varphi}^{2k}\left(1-\frac{\widetilde{\varphi}^2}{6l}\right)^{2k}
\exp\left(\frac{2im\widetilde{\varphi}}{\sqrt{l}}\right)\exp\left(-\left(2l+2n_i-2k\right)\left(\frac{\widetilde{\varphi}^2}{2l}+\frac{\widetilde{\varphi}^4}{12l^2}\right)\right)\\
\sim & \frac{1}{l^{k+\frac{1}{2}}}\int_{-\infty}^{\infty} d\widetilde{\varphi} \widetilde{\varphi}^{2k} \left(1+ \mathcal{B}_{i,2}\right)\exp\left(-\widetilde{\varphi}^2+\frac{2im\widetilde{\varphi}}{\sqrt{l}}+\mathcal{A}_{i,1}+\mathcal{A}_{i,2}\right).
\end{align*}Here and in the following, we denote by $\mathcal{X}_{i,1}$ and $\mathcal{X}_{i,2}$ terms of order $\sqrt{\vep}$ and $\vep$ respectively. Similarly, we have
\begin{align*}
&\int_{-\frac{\pi}{2}}^{\frac{\pi}{2}}d\varphi'  \cos^{ 2j_i'-2k'}\varphi' \sin^{2k'}\varphi' e^{2im\varphi'}\\
\sim &\frac{1}{l^{k'+\frac{1}{2}}}\int_{-\infty}^{\infty} d\widetilde{\varphi}' \widetilde{\varphi}^{\prime 2k'}\left(1-\frac{\widetilde{\varphi}^{\prime 2}}{6l}\right)^{2k'}
\exp\left(\frac{2im\widetilde{\varphi}'}{\sqrt{l}}\right)\exp\left(-\left(\frac{2b}{a}l+\frac{b}{a}(n_i+n_{i+1})+2q_i-2k'\right)
\left(\frac{\widetilde{\varphi}^{\prime 2}}{2l}+\frac{\widetilde{\varphi}^{\prime 4}}{12l^2}\right)\right)\\
\sim & \frac{1}{l^{k'+\frac{1}{2}}}\int_{-\infty}^{\infty} d\widetilde{\varphi}' \widetilde{\varphi}^{\prime 2k} \left(1+ \mathcal{D}_{i,2}\right)\exp\left(-\frac{b}{a}\widetilde{\varphi}^{\prime 2}+\frac{2im\widetilde{\varphi}'}{\sqrt{l}}+\mathcal{C}_{i,1}+\mathcal{C}_{i,2}\right).
\end{align*}
On the other hand, making a change of variables
$$\theta=\widetilde{\theta}+\theta_0,$$where
\begin{align*}
\sinh\theta_0=\frac{\tau}{\sqrt{1-\tau^2}},
\end{align*}
we have expansions of the form
\begin{align*}
A\sim &-l^{\frac{3}{2}}\sqrt{\frac{1-\tau}{1+\tau}}\frac{2ib}{a}\left(\widetilde{\varphi}+\widetilde{\varphi}'\right), \\
B\sim &\frac{2b}{a}l^2\left(1+\mathcal{E}_{i,1}+\mathcal{E}_{i,2}\right),\\
C\sim &-\frac{2b}{a}l^2\left(1+\mathcal{F}_{i,1}+\mathcal{F}_{i,2}\right),\\
D\sim &l^{2}\sqrt{\frac{1+\tau}{1-\tau}}\frac{2b}{a}\left(\frac{i\left(\widetilde{\varphi}+\widetilde{\varphi}'\right)}{\sqrt{l}}+\frac{m}{bl}\right),\\
\end{align*}  and
\begin{align*}
&\int_{0}^{\infty} d\theta\sinh\theta
 e^{ -\kappa L\cosh\theta} \exp\left(\left[\frac{l}{a}+\frac{a+1}{2a}n_i+\frac{b}{2a}n_{i+1}+q_i-2k-2k'\right]\theta\right)\\
 \sim &\int_{-\infty}^{\infty} d\widetilde{\theta}\sinh\left(\widetilde{\theta}+\theta_0\right)
 e^{ -(1+\vep)\omega\cosh\left(\widetilde{\theta}+\theta_0\right)} \exp\left(\left[\frac{l}{a}+\frac{a+1}{2a}n_i+\frac{b}{2a}n_{i+1}+q_i-2k-2k'\right]\left(\widetilde{\theta}+\theta_0\right)\right)\\
 \sim &\frac{\tau}{\sqrt{1-\tau^2}}\left(\frac{1+\tau}{1-\tau}\right)^{\frac{l}{2a}+\frac{a+1}{4a}n_i+\frac{b}{4a}n_{i+1}+\frac{q_i}{2}-k-k'}
  \int_{-\infty}^{\infty} d\widetilde{\theta}\exp\left(-\frac{l}{a\tau}-\frac{l}{2a\tau}\widetilde{\theta}^2-\frac{\vep l}{a\tau}
  +\left(\frac{a+1}{2a}n_i+\frac{b}{2a}n_{i+1}+q_i\right)\widetilde{\theta}\right)\\&\times
  \left(1+\mathcal{G}_{i,1}+\mathcal{G}_{i,2}\right)\exp\left(\mathcal{H}_{i,1}+\mathcal{H}_{i,2}\right).
\end{align*}Using Stirling's asymptotic expansion for gamma functions, we also obtain an asymptotic expansion
\begin{align*}
 &\frac{1}{\pi^2}\frac{\displaystyle\sqrt{ (j_i-m)!\left(j_i'-m\right)! (j_i+m)!\left(j_i'+m\right)!}}{\displaystyle\left(j_i+\frac{1}{2}-k\right)!\left(j_i'+\frac{1}{2}-k'\right)!}\\
 \sim &\frac{1}{\pi^2} \left(\frac{b}{a}\right)^{k'-\frac{1}{2}}l^{k+k'-1}\exp\left(\frac{m^2}{2bl}+\mathcal{I}_{i,1}+\mathcal{I}_{i,2}\right).
\end{align*}
Collecting the terms, we have
\begin{align*}
\mathbb{V}^{1}_{j_im, j_i'm}\sim & \frac{2}{\pi^2} \left(\frac{b}{a}\right)^{k'+\frac{1}{2}}\frac{\tau}{\sqrt{1-\tau^2}}\left(\frac{1+\tau}{1-\tau}\right)^{\frac{l}{2a}+\frac{a+1}{4a}n_i+\frac{b}{4a}n_{i+1}+\frac{q_i}{2}-k-k'}
\sum_{k=0}^{\infty}\sum_{k'=0}^{\infty}\frac{1}{k!k'!}\\&\times \int_{-\infty}^{\infty} d\widetilde{\varphi} \widetilde{\varphi}^{2k} \left(1+ \mathcal{B}_{i,2}\right)\exp\left(-\widetilde{\varphi}^2+\frac{2im\widetilde{\varphi}}{\sqrt{l}}+\mathcal{A}_{i,1}+\mathcal{A}_{i,2}\right)\\&\times\int_{-\infty}^{\infty} d\widetilde{\varphi}' \widetilde{\varphi}^{\prime 2k} \left(1+ \mathcal{D}_{i,2}\right)\exp\left(-\frac{b}{a}\widetilde{\varphi}^{\prime 2}+\frac{2im\widetilde{\varphi}'}{\sqrt{l}}+\mathcal{C}_{i,1}+\mathcal{C}_{i,2}\right)\\
&\times  \int_{-\infty}^{\infty} d\widetilde{\theta}\exp\left(-\frac{l}{a\tau}-\frac{l}{2a\tau}\widetilde{\theta}^2-\frac{\vep l}{a\tau}
  +\left(\frac{a+1}{2a}n_i+\frac{b}{2a}n_{i+1}+q_i\right)\widetilde{\theta}\right)\\&\times
  \left(1+\mathcal{G}_{i,1}+\mathcal{G}_{i,2}\right)\exp\left(\mathcal{H}_{i,1}+\mathcal{H}_{i,2}\right)\exp\left(\frac{m^2}{2bl}+\mathcal{I}_{i,1}+\mathcal{I}_{i,2}\right)\\
  &\times\begin{pmatrix} \displaystyle- \sqrt{\frac{1-\tau}{1+\tau}}\frac{i(\widetilde{\varphi}+\widetilde{\varphi}')}{\sqrt{l}}&1+\mathcal{E}_{i,1}+\mathcal{E}_{i,2} \\
  -\left(1+\mathcal{F}_{i,1}+\mathcal{F}_{i,2}\right)&\displaystyle \sqrt{\frac{1+\tau}{1-\tau}}\left(\frac{i\left(\widetilde{\varphi}+\widetilde{\varphi}'\right)}{\sqrt{l}}+\frac{m}{bl}\right)\end{pmatrix}.\end{align*}
  Now we expand the terms to obtain terms of order $\sqrt{\vep}$ and $\vep$ respectively, and then perform the summation over $k$ and $k'$. After summation, the integrations over $\widetilde{\varphi}$ and $\widetilde{\varphi}'$ are standard, followed by the integration over $\widetilde{\theta}$. We obtain an expression of the form
  \begin{align*}  \mathbb{V}^{1}_{j_im, j_i'm}  \sim  & \sqrt{\frac{2a\tau}{\pi l}}  \left(\frac{1+\tau}{1-\tau}\right)^{\frac{l}{2a}+\frac{a+1}{4a}n_i+\frac{b}{4a}n_{i+1}+\frac{q_i}{2}+\frac{1}{2}}
     \exp\left(-\frac{l}{a\tau} -\frac{\vep l}{a\tau}
  +\frac{a\tau}{2l}\left(\frac{a+1}{2a}n_i+\frac{b}{2a}n_{i+1}+q_i\right)^2-\frac{m^2}{2bl\tau}\right)
  \\
  &\times\left(1+\mathcal{J}_{i,1}+\mathcal{J}_{i,2}\right)\begin{pmatrix} \displaystyle \sqrt{1-\tau^2}\frac{m}{2lb\tau}&1+\mathcal{K}_{i,1}+\mathcal{K}_{i,2} \\
  -\left(1+\mathcal{K}_{i,1}+\mathcal{K}_{i,2}\right)&\displaystyle -\sqrt{1-\tau^2}\frac{m}{2lb\tau}\end{pmatrix}.
\end{align*}
Interchanging $n_i$ and $n_{i+1}$ gives
\begin{align*}
\mathbb{V}^{2}_{j_i'm, j_{i+1}m}\sim & \sqrt{\frac{2a\tau}{\pi l}}  \left(\frac{1+\tau}{1-\tau}\right)^{\frac{l}{2a}+\frac{a+1}{4a}n_{i+1}+\frac{b}{4a}n_{i}+\frac{q_i}{2}+\frac{1}{2}}
     \exp\left(-\frac{l}{a\tau} -\frac{\vep l}{a\tau}
  +\frac{a\tau}{2l}\left(\frac{a+1}{2a}n_{i+1}+\frac{b}{2a}n_{i}+q_i\right)^2-\frac{m^2}{2bl\tau}\right)
  \\
  &\times\left(1+\mathcal{L}_{i,1}+\mathcal{L}_{i,2}\right)\begin{pmatrix} \displaystyle\sqrt{1-\tau^2}\frac{m}{2lb\tau}&1+\mathcal{M}_{i,1}+\mathcal{M}_{i,2} \\
  -\left(1+\mathcal{M}_{i,1}+\mathcal{M}_{i,2}\right)&\displaystyle -\sqrt{1-\tau^2}\frac{m}{2lb\tau}\end{pmatrix}.
\end{align*}Now for $T_{j_i'm,1}^{(+), B}$ and $T_{j_i'm,2}^{(+), B}$,
\begin{align*}
T_{j_i'm,1}^{(+), B}=\overline{T_{j_i'm,2}^{(+), B}}=&\frac{I_{j_i'}(b\omega)- iI_{j_i'+1}(b\omega)}{K_{j_i'}(b\omega)+ iK_{j_i'+1}(b\omega)}\\=&\frac{I_{j_i'}(b\omega)}{K_{j_i'}(b\omega)}\frac{\displaystyle 1
- i\frac{I_{j_i'+1}(b\omega)}{I_{j_i'}(b\omega)}}{\displaystyle 1+ i\frac{K_{j_i'+1}(b\omega)}{K_{j_i'}(b\omega)}},
\end{align*}
we use Debye asymptotic expansions of modified Bessel functions
\begin{align*}
I_{\nu}(\nu z)\sim & \frac{1}{\sqrt{2\pi \nu}}\frac{e^{\nu\eta(z)}}{(1+z^2)^{\frac{1}{4}}}\left(1+\frac{u_1(t(z))}{\nu}+\ldots\right),\\
K_{\nu}(\nu z)\sim &\sqrt{\frac{\pi}{ 2 \nu}}\frac{e^{-\nu\eta(z)}}{(1+z^2)^{\frac{1}{4}}}\left(1-\frac{u_1(t(z))}{\nu}+\ldots\right),
\end{align*}where
\begin{gather*}
\eta(z)=\sqrt{1+z^2}+\log\frac{z}{1+\sqrt{1+z^2}}\\ t(z)=\frac{1}{\sqrt{1+z^2}},\quad u_1(t)=\frac{t}{8}-\frac{5t^3}{24}.
\end{gather*}We find that
\begin{align*}
\frac{I_{j_i'}(b\omega)}{K_{j_i'}(b\omega)}
\sim & \frac{1}{\pi}\left(\frac{1+\tau}{1-\tau}\right)^{-\frac{b}{2a}(n_i+n_{i+1})-q_i-\frac{bl}{a}}\exp\left(\frac{2bl}{a\tau}
-\frac{a\tau}{ bl}\left(\frac{b}{2a}(n_i+n_{i+1})+ q_i\right)^2+\mathcal{N}_{i,1}+\mathcal{N}_{i,2}\right) \left(1+\mathcal{O}_{2}\right),
\end{align*}
and
\begin{align*}
\frac{\displaystyle 1
- i\frac{I_{j_i'+1}(b\omega)}{I_{j_i'}(b\omega)}}{\displaystyle 1+ i\frac{K_{j_i'+1}(b\omega)}{K_{j_i'}(b\omega)}}
\sim & -i\sqrt{\frac{1-\tau}{1+\tau}}\Bigl(1+\mathcal{P}_{i,1}+ \mathcal{P}_{i,2} \Bigr).
\end{align*}Here $\mathcal{P}_{i,1}$ is real and $\mathcal{P}_{i,2}$ is complex.
After expansion and simplification, we find that
\begin{align*}
&\int_{-\infty}^{\infty} dq_i\mathbb{V}^1_{j_im,j_i'm}
 \begin{pmatrix} T_{j_i'm,1}^{+, B}&0\\0&T_{j_i'm, 2}^{+, B}\end{pmatrix}\mathbb{V}^2_{j_i'm, j_{i+1}m}\\
 \sim &-i\frac{2a\tau}{\pi^2 l}\left(\frac{1+\tau}{1-\tau}\right)^{l+\frac{n_i+n_{i+1}}{2}+\frac{1}{2}}\exp\left(-\frac{2l}{\tau}-\frac{2\vep l}{a\tau}-\frac{m^2}{bl\tau}
 +\frac{\tau(1+a)}{4l}(n_i^2+n_{i+1}^2)+\frac{\tau(1-a)}{2l}n_in_{i+1}\right)\\
 &\times \int_{-\infty}^{\infty} dq_i\exp\left(-\frac{a^2\tau}{bl}q_i^2\right)\left(1+\mathcal{Q}_{i,1}+\mathcal{Q}_{i,2}\right) \left(1+\mathcal{O}_2\right)  \\&\times\begin{pmatrix} \displaystyle 1 +\overline{\mathcal{P}_{i,2}}+\left(\sqrt{1-\tau^2}\frac{m}{2lb\tau}\right)^2& \displaystyle\sqrt{1-\tau^2}\frac{m}{lb\tau} \\
  -\displaystyle\sqrt{1-\tau^2}\frac{m}{lb\tau}&\displaystyle -\left(1 +\mathcal{P}_{i,2} \right)-\left(\sqrt{1-\tau^2}\frac{m}{2lb\tau}\right)^2\end{pmatrix}\\
  =&-i\frac{2 }{\pi^{\frac{3}{2}}}\sqrt{\frac{b\tau}{l}}\left(\frac{1+\tau}{1-\tau}\right)^{l+\frac{n_i+n_{i+1}}{2}+\frac{1}{2}}\exp\left(-\frac{2l}{\tau}-\frac{2\vep l}{a\tau}-\frac{m^2}{bl\tau}
 +\frac{\tau(1+a)}{4l}(n_i^2+n_{i+1}^2)+\frac{\tau(1-a)}{2l}n_in_{i+1}\right)\\
 &\times  \left(1+\mathcal{R}_{i,1}+\mathcal{R}_{i,2}\right) \left(1+\mathcal{O}_2\right) \begin{pmatrix} \displaystyle 1 +\overline{\mathcal{S}_{i,2}}+\left(\sqrt{1-\tau^2}\frac{m}{2lb\tau}\right)^2& \displaystyle\sqrt{1-\tau^2}\frac{m}{lb\tau} \\
  -\displaystyle\sqrt{1-\tau^2}\frac{m}{lb\tau}&\displaystyle -\left(1 +\mathcal{S}_{i,2} \right)-\left(\sqrt{1-\tau^2}\frac{m}{2lb\tau}\right)^2\end{pmatrix}.
\end{align*}For $T_{j_im,1}^{(+), A}$ and $T_{j_im,2}^{(+), A}$, we have similar expansions
\begin{align*}
\frac{I_{j_i}(a\omega)}{K_{j_i}(a\omega)}\sim &  \frac{1}{\pi}\left(\frac{1+\tau}{1-\tau}\right)^{-\frac{1}{2 }(n_i-n_{i+1})-\frac{1}{2 }(n_i+n_{i+1})- l}\exp\left(\frac{2l}{\tau}
-\frac{\tau}{ l} n_i^2+\mathcal{T}_{i,1}+\mathcal{T}_{i,2}\right) \left(1+\mathcal{U}_{2}\right),
\end{align*}
and
\begin{align*}
\frac{\displaystyle 1
- i\frac{I_{j_i+1}(a\omega)}{I_{j_i}(a\omega)}}{\displaystyle 1+ i\frac{K_{j_i+1}(a\omega)}{K_{j_i}(a\omega)}}
\sim & -i\sqrt{\frac{1-\tau}{1+\tau}}\Bigl(1+\mathcal{V}_{i,1}+ \mathcal{V}_{i,2}  \Bigr).
\end{align*}Finally,
we obtain an expansion of the form
\begin{align*}
\mathbb{M}_{j_im, j_{i+1}m}\sim &  - \frac{1 }{2\sqrt{\pi}}\sqrt{\frac{b\tau}{l}}C^{ n_i-n_{i+1} }\exp\left( -\frac{2\vep l}{a\tau}-\frac{m^2}{bl\tau}
-\frac{b\tau }{4l}(n_i-n_{i+1})^2 \right)  \\
 &\times  \left(1+\mathcal{W}_{i,1}+\mathcal{W}_{i,2}\right) \left(1+\mathcal{O}_{2}+\mathcal{U}_2\right)   \begin{pmatrix} \displaystyle 1 & \displaystyle\sqrt{1-\tau^2}\frac{m}{lb\tau} \\
  \displaystyle\sqrt{1-\tau^2}\frac{m}{lb\tau}& 1\end{pmatrix}.
\end{align*}The imaginary part has been omitted since it is easy to verify that it would not contribute in the trace.

Substituting into \eqref{eq02_12_2}, we find that the term of order $\sqrt{\vep}$ is zero and we have
\begin{align*}
E_{\text{Cas}}=&\frac{2\hbar c b^{\frac{s+1}{2}}  }{2^{s+1}\pi^{\frac{s+3}{2}} r_A}\sum_{s=0}^{\infty}\frac{(-1)^{s+1}}{s+1}\int_0^{1}d\tau \frac{\tau^{\frac{s-3}{2}}}{ \sqrt{1-\tau^2}} \int_0^{\infty} dl \,l^{-\frac{s-1}{2}}
\int_{-\infty}^{\infty} dm \int_{-\infty}^{\infty} dn_1\ldots\int_{-\infty}^{\infty} dn_s\\&\times \exp\left( -\frac{2\vep(s+1) l}{a\tau}-\frac{(s+1)m^2}{bl\tau}
-\frac{b\tau }{4l}\sum_{i=0}^s(n_i-n_{i+1})^2 \right) \\&\times\left(1+\sum_{i=0}^{s-1}\sum_{j=i+1}^s\mathcal{W}_{i,1}\mathcal{W}_{j,1}+\sum_{i=0}^s\mathcal{W}_{i,2}+\frac{s(s+1)}{2}\left(\sqrt{1-\tau^2}\frac{m}{lb\tau}\right)^2
 +(s+1)(\mathscr{B}_{2}+\mathcal{U}_2)\right).
\end{align*}The integrations over $n_i$, $m$, $l$ and $\tau$ are standard and we obtain
\begin{align*}
E_{\text{Cas}}\sim & \frac{ \hbar c ba^2 }{4 \pi \vep^2 r_A}\sum_{s=0}^{\infty}\frac{(-1)^{s+1}}{(s+1)^{4}}\left(1+\vep\left(-1+\frac{1-(s+1)^2}{3ab}\right)+\ldots\right)\\
  =&-\frac{ 7\pi^3\hbar c r_Ar_B }{2880  d^2 (r_A+r_B)} \left(1-\frac{d}{r_A+r_B}+\left[\frac{1}{3}-\frac{20}{7\pi^2}\right]\left(\frac{d}{r_A}+\frac{d}{r_B}\right)+\ldots\right).
\end{align*}One can easily verify that the leading order term
\begin{align*}
E_{\text{Cas}}\sim &-\frac{ 7\hbar c r_Ar_B }{2880 \pi d^2 (r_A+r_B)}
\end{align*}coincides with the proximity force approximation. Since
$$\frac{1}{3}-\frac{20}{7\pi^2}=0.0438,$$the sign of the next-to-leading order term can be the same as  or different from the leading term depending on the ratio of $r_A$ to $r_B$.

We can also compare this result to the results of scalar fields and electromagnetic fields \cite{13}:
\begin{align*}
E_{\text{Cas}}^{\text{D}}\sim & -\frac{  \pi^3\hbar c r_Ar_B }{1440  d^2 (r_A+r_B)} \left(1-\frac{d}{r_A+r_B}+ \frac{1}{3} \left(\frac{d}{r_A}+\frac{d}{r_B}\right)+\ldots\right),\\
E_{\text{Cas}}^{\text{N}}\sim & -\frac{  \pi^3\hbar c r_Ar_B }{1440  d^2 (r_A+r_B)} \left(1-\frac{d}{r_A+r_B}+ \left[\frac{1}{3}-\frac{40}{\pi^2}\right] \left(\frac{d}{r_A}+\frac{d}{r_B}\right)+\ldots\right),\\
E_{\text{Cas}}^{\text{C}}\sim & -\frac{  \pi^3\hbar c r_Ar_B }{720  d^2 (r_A+r_B)} \left(1-\frac{d}{r_A+r_B}+ \left[\frac{1}{3}-\frac{20}{\pi^2}\right] \left(\frac{d}{r_A}+\frac{d}{r_B}\right)+\ldots\right).
\end{align*}
For the case of two Dirichlet spheres, the sign of the next-to-leading order term can also be the same as  or different from the leading term depending on the ratio of $r_A$ to $r_B$.  However, for the case of two Neumann spheres or two perfectly conducting spheres, the sign of the next-to-leading order term is always different from the leading term.

It is also interesting to note that for scalar field, fermionic field as well as the electromagnetic field, the ratio of the next-to-leading order term to the leading order term  always contain the term
$$-\frac{d}{r_A+r_B}$$with same coefficient. It looks like this is a universal term that does not depend on boundary conditions.

\section{Conclusion}
In this work, we consider the Casimir interaction between two spheres that results from the vacuum fluctuations of a massless Dirac  field with MIT-bag boundary conditions. The sphere-sphere Casimir interaction due to scalar fields and electromagnetic fields has been well-understood. However, to the best of our knowledge, we are the first one that investigates the fermionic interaction between two spheres. Using our formalism in \cite{1}, we derive the functional representation of the Casimir interaction energy. The most technical part in the derivation is the computation of the translation matrix  that relates  the fermionic spherical waves in two different coordinate systems. We tackle the problem using the operator approach as in \cite{4,1}. The result can be considered as an important byproduct of this work.

From the formula of the Casimir interaction energy, we compute the large separation and small separation asymptotic behaviors. As usual, the large separation asymptotic behavior is very easy to compute since it only involves a few terms that can be computed explicitly. We showed that when the separation between the spheres is large, the Casimir interaction energy behaves like
$$E_{\text{Cas}}\sim -\frac{6\hbar c r_A^2 r_B^2}{\pi L^5},$$
where $L$ is the separation between the centers of the two spheres, and $r_A$ and $r_B$ are the radii of the spheres. We note that the order of the Casimir interaction energy is $L^{-5}$, which is intermediate between the order of the Casimir interaction energy between two Dirichlet spheres ($L^{-3}$) and the order of the Casimir interaction energy between two Neumann spheres ($L^{-7}$).

In the small separation limit, we compute analytically the asymptotic behavior of the Casimir interaction energy up to the next-to-leading order term and find that
\begin{align*}
E_{\text{Cas}}\sim &  -\frac{ 7\pi^3\hbar c r_Ar_B }{2880  d^2 (r_A+r_B)} \left(1-\frac{d}{r_A+r_B}+\left[\frac{1}{3}-\frac{20}{7\pi^2}\right]\left(\frac{d}{r_A}+\frac{d}{r_B}\right)+\ldots\right).
\end{align*}Here $d$ is the distance between the two spheres. The leading order term agrees with the proximity force approximation. We also note that the term $$-\frac{d}{r_A+r_B}$$ which appear in the ratio of the next-to-leading order term to the leading order term is 'universal' among different types of quantum field.

From the leading terms of the small separation and large separation asymptotic behaviors, we find that the Casimir interaction force is an attractive force in these two regimes. We believe that the force is always attractive at all separations, but it is not easy to deduce this from the expression for the Casimir interaction energy.

\bigskip
\begin{acknowledgments}\noindent
  This work is supported by the Ministry of Higher Education of Malaysia  under the  FRGS grant FRGS/1/2013/ST02/UNIM/02/2.
\end{acknowledgments}

\end{document}